%% file: IEEE-conference-template-062824.tex
\documentclass[conference]{IEEEtran}

\usepackage{cite}
\usepackage{amsmath,amssymb,amsfonts}
\usepackage{algorithm}
\usepackage{algorithmic}
\usepackage{graphicx}
\usepackage{textcomp}
\usepackage{xcolor}
\usepackage{booktabs}
\usepackage{url}
\usepackage{pdfpages}

\usepackage{cite}
\usepackage{amsmath,amssymb,amsfonts}
\usepackage{graphicx}
\usepackage{textcomp}
\usepackage{xcolor}
\usepackage{booktabs}
\usepackage{url}
\usepackage{sc26repro}

\def\BibTeX{{\rm B\kern-.05em{\sc i\kern-.025em b}\kern-.08em
    T\kern-.1667em\lower.7ex\hbox{E}\kern-.125emX}}
\begin{document}

\title{Analytical Resource Management for Fine-grained MoE Computation-Communication Overlap}


\author{\IEEEauthorblockN{Hongyu Liu}
\IEEEauthorblockA{
\textit{Chalmers University of Technology}\\
\textit{and University of Gothenburg}\\
Gothenburg, Sweden \\ 
hongyuli@chalmers.se}
\and
\IEEEauthorblockN{Minyu Cui}
\IEEEauthorblockA{
\textit{Linnaeus University}\\
Växjö, Sweden \\
minyu.cui@lnu.se}
\and
\IEEEauthorblockN{Miquel Peric\`as}
\IEEEauthorblockA{
\textit{Chalmers University of Technology}\\
\textit{and University of Gothenburg}\\
Gothenburg, Sweden \\ 
miquelp@chalmers.se}
}

\maketitle

\begin{abstract}
Fine-grained computation--communication overlap in distributed
Mixture-of-Experts (MoE) inference allows communication to begin as partial compute results become ready. 
%
However, cooperative thread arrays (CTAs) performing computation and communication contend for finite residency capacity on the streaming multiprocessors (SMs). Because a resident CTA generally retains its allocated SM resources until completion, CTAs that cannot be co-resident must wait for resources to become available, resulting in wave-like execution.
A fixed resource partition cannot adapt to changes in input size, routed expert load, and kernel
configuration, potentially causing a communication backlog or reducing expert
compute parallelism. 

We present a wave-quantized analytical model and
launch-time resource manager for dependency-coupled overlap pipelines. Using
the current routed-tile counts, kernel occupancy, GPU residency constraints,
and split-level readiness dependencies, it selects the communication-CTA count
and resource partition before each kernel launch without candidate execution,
per-workload profiling, or kernel recompilation. We integrate the method into
the public COMET A100 implementation in the FLUX codebase.
We evaluate three MoE models on four NVIDIA A100 GPUs under 
several parallelism strategies
at the GEMM2+GatherRS operator,
complete post-router MoE layer, and complete-model prefill levels.
Across 15 real-p90 workloads, the analytical selector achieves 3.22\% mean
regret relative to the measured oracle with a mean solver overhead of
0.157~\(\mu\)s.
Over COMET, our method achieves geometric-mean speedups of
\(2.528\times\) at the GEMM2+GatherRS operator, with a maximum of
\(4.218\times\); \(1.771\times\) at the complete post-router MoE layer,
with a maximum of \(2.584\times\); and \(1.185\times\) for
complete-model prefill, with a maximum of \(1.439\times\).
For the representative TP=2/EP=2 configuration, complete-model
prefill achieves 
geometric-mean 
speedups over COMET of \(1.336\times\), \(1.211\times\), and
\(1.232\times\) for Granite, Qwen, and DeepSeek-V2-Lite, respectively, with a best
speedup of \(1.439\times\). At every feasible TP=2/EP=2 sequence length of at least
4,096, our implementation outperforms COMET and the other state-of-the-art MoE
systems, Megatron core-TE and FastMoE TP+NCCL.


\end{abstract}

\begin{IEEEkeywords}
Mixture-of-Experts, computation--communication overlap, GPU resource
management, analytical performance modeling
\end{IEEEkeywords}

\section{Introduction}\label{Intro}




Recent advances in large language models (LLMs) have been driven in part by
scaling model parameters and training data. Scaling dense models, however,
increases both memory consumption and the computation performed for every
token, causing training and inference costs to grow rapidly.
Mixture-of-Experts (MoE) models offer a more scalable alternative: they increase parameter capacity by incorporating many experts while sparsely activating only a small subset for each token, thereby avoiding a proportional increase in computational overhead.

As models and expert sets continue to grow, a single GPU can no longer store
or efficiently execute all experts. Distributed MoE inference must place
experts across multiple GPUs, send routed tokens to the devices that host
their selected experts, and return the expert outputs to the devices that own
the original tokens. Prior studies report that this cross-device
communication can account for as much as 47\% of end-to-end model execution
time in representative models and frameworks~\cite{comet}.
Communication thus limits not only an individual MoE layer but also the
scaling benefit expected from adding GPUs. Overlapping this communication 
with expert GEMMs without degrading their efficiency is therefore critical to 
scaling distributed MoE inference.


A representative solution~\cite{comet} is
\emph{dependency-aware, thread-block-specialized fine-grained overlap}. 
This technique exposes result readiness at fine granularity, 
allowing a communication CTA to process a local result as soon as it
becomes available rather than waiting for the entire GEMM to finish. 
Implementations may assign compute and communication work to separate CTAs, either in 
concurrently executing kernels or within a fused kernel. 
These role-specialized CTAs keep long-latency communication instructions out
of compute CTAs, thereby avoiding direct instruction-level interference with
computation.
However, both CTA classes compete for limited SM residency and shared GPU
resources. This introduces a resource-partitioning problem that ultimately
determines the effectiveness of overlap. With insufficient communication
resources, ready results cannot be transferred and reduced promptly, leaving
a communication tail on the critical path. With excessive communication
resources, expert-GEMM parallelism decreases and computation becomes the
bottleneck. Therefore, effective overlap requires resource management to
balance the two CTA classes under dependency constraints to minimize the
overall pipeline latency, rather than maximizing either throughput in isolation.
An imbalanced allocation can create backpressure or resource contention and may perform worse than a non-overlapped schedule.

No single communication-CTA allocation is optimal for every workload.
Token count, expert matrix dimensions, the number of routed expert tiles, and the pipeline split granularity change computation and communication work at
different rates. Moreover, GPUs execute integer numbers of CTAs and tile
waves, so performance changes discontinuously at wave boundaries instead of scaling smoothly with a resource ratio. A fixed partition cannot adapt to
this shape-dependent behavior. Profiling candidate configurations for every shape can identify an efficient partition, but it incurs deployment and tuning costs
and does not provide a principled solution for an unseen workload. This leads to our
central question: given the workload shape, dependency structure, and kernel
resource requirements before execution, can we directly derive the resource partition between
computation and communication CTAs without per-shape profiling?
%
The key difficulty is \emph{dependency--residency coupling}: communication CTAs
can advance only after their inputs become ready, yet their residency reduces
the SM capacity available to compute CTAs. Because both sides progress in
discrete CTA waves, the optimal partition minimizes overall pipeline latency
rather than either side's isolated throughput.

Based on this observation, we formulate resource management for
thread-block-specialized overlap as \emph{dependency-constrained resource
partitioning} and develop a \emph{wave-quantized analytical makespan model}.
Rather than adding new precompiled kernel variants, we insert an
\emph{analytical launch-time resource manager} between conventional kernel
dispatch and the actual launch. 
This design decouples kernel implementation from
workload-specific resource allocation: the same binaries adapt at runtime
without recompilation, candidate-kernel execution, warm-up timing, or
consulting a per-workload winner table. The core analytical decision takes
approximately 0.16~$\mu$s and therefore incurs negligible overhead.
%
The contributions of this paper are threefold:
\begin{itemize}
\item \textbf{Problem characterization and empirical finding.}
We characterize CTA allocation in dependency-coupled,
thread-block-specialized overlap as a non-preemptive GPU-residency
partitioning problem, and identify the dependency-residency coupling through
which communication progress and compute capacity jointly determine pipeline
performance.

\item \textbf{Analytical model and launch-time resource management.}
We develop a dependency-constrained, wave-quantized makespan model and
implement a lightweight runtime resource manager between precompiled kernel
dispatch and launch. Using the current workload, actual routed tile counts,
kernel residency, and dependencies, it directly derives communication CTA count \(C\), determines the associated
communication reservation \(R(C)\), 
and computes the remaining compute capacity
\(P(C)\).

\item \textbf{System implementation and cross-model, multi-level evaluation.}
We integrate this analytical launch-time resource manager
into COMET's public A100 implementation in the FLUX codebase~\cite{flux,comet}
and evaluate its accuracy and performance benefits at the
three measurement levels: the GEMM2+GatherRS operator, the complete post-router
MoE layer, and complete-model prefill. The reported speedups arise solely from improved CTA
resource allocation: our method does not reduce FLOPs, routed tokens, or
communication volume, and it preserves numerical correctness.
At the first two levels, we evaluate
Granite-3.1-1B-A400M, Qwen1.5-MoE-A2.7B, and DeepSeek-V2-Lite under
uniform, real-p50, and real-p90 routing workloads. At the complete-model
prefill level, we compare our system with upstream COMET,
Megatron core-TE, and FastMoE TP+NCCL using model-matched inputs and identical execution and measurement boundaries.
\end{itemize}

\input{background}

\section{Dependency-Constrained, Wave-Quantized Resource Management}
\label{Method}


This section presents our dependency-constrained, wave-quantized model and
its runtime resource manager.

\subsection{Optimization Scope}

\textbf{Optimization scope.}
Across the TP=2/EP=2 evaluation matrix, the ratio of baseline standalone
GEMM2+GatherRS latency to complete post-router MoE layer latency has a
geometric mean of 78.2\%. The corresponding geometric means are 85.3\%,
75.6\%, and 74.2\% for Granite, Qwen, and DeepSeek-V2-Lite, respectively.
Thus, this operator is a major component of the complete post-router MoE layer.

\begin{figure*}[t]
\centering
\includegraphics[width=\textwidth]{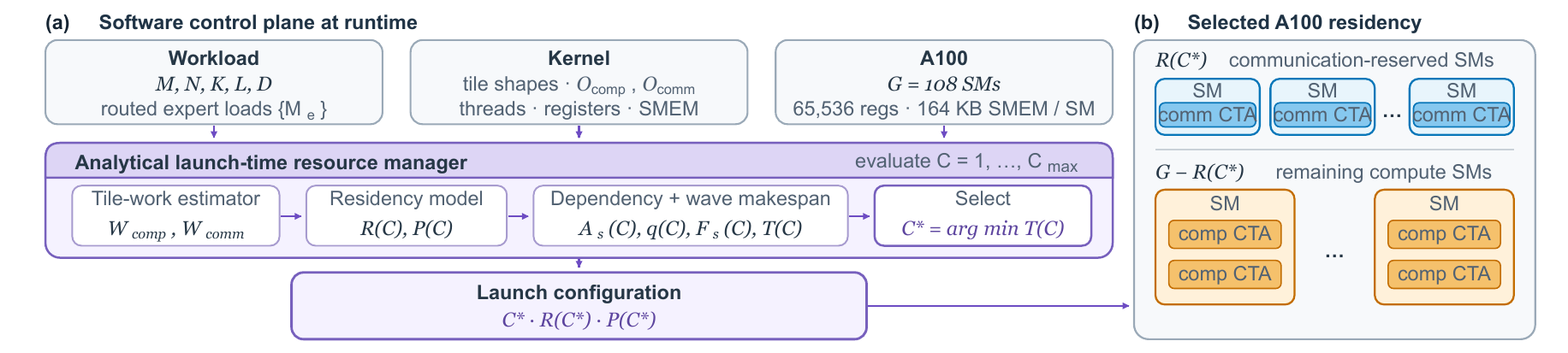}
\caption{Launch-time resource selection and A100 residency.
(a) The analytical manager selects \(C^\ast\), \(R(C^\ast)\), and
\(P(C^\ast)\) from workload, kernel, and GPU inputs.
(b) The resulting compute and communication residency.}
\label{fig:launch-time-resource-control}
\end{figure*}

It also exhibits a clear resource mismatch. A fixed GatherRS communication-CTA
count cannot adapt to changes in token count, matrix dimensions, split count,
or the number of GEMM2 tiles, and can produce a suboptimal tradeoff between
insufficient communication parallelism and reduced GEMM2 capacity. In
contrast, AllGather+GEMM1 already launches computation incrementally from AllGather
readiness and does not exhibit the same fixed-CTA partitioning bottleneck. We
therefore optimize the CTA resource partition of GEMM2+GatherRS while leaving
AllGather+GEMM1 and the activation unchanged.
%
%
%
Fig.~\ref{fig:launch-time-resource-control} summarizes the control path from
per-invocation metadata to the selected launch configuration and its logical
CTA-to-SM residency. The following subsections define the modeled quantities
and selection procedure.

\subsection{Dependency-Constrained Resource Partitioning}

\textbf{Resource notation.}
Table~\ref{tab:resource-notation} summarizes the static relationship between
the GPU and the two kernel roles.

\begin{table}[!b]
\centering
\caption{Resource notation for the GEMM2+GatherRS operator.}
\label{tab:resource-notation}
\footnotesize
\setlength{\tabcolsep}{3pt}
\renewcommand{\arraystretch}{1.08}
\begin{tabular}{@{}lp{0.76\columnwidth}@{}}
\toprule
\textbf{Symbol} & \textbf{Definition} \\
\midrule
\(G\) & SMs available to the GEMM2+GatherRS operator; 108 on the evaluated A100. \\
\(C\) & GatherRS communication CTAs launched; the primary decision variable. \\
\(o_{\mathrm{comm}}\) & Maximum resident communication CTAs per SM. \\
\(o_{\mathrm{comp}}\) & Maximum resident GEMM2 compute CTAs per SM. \\
\(R(C)\) & Minimum SM residency positions reserved for \(C\) communication CTAs. \\
\(P(C)\) & Resident compute-CTA slots remaining after the reservation. \\
\bottomrule
\end{tabular}
\end{table}

\textbf{Minimum progress-preserving reservation.}
For all \(C\) communication CTAs to acquire residency and advance the
dependency pipeline, the minimum reservation is
\begin{equation}
R(C)=\left\lceil\frac{C}{o_{\mathrm{comm}}}\right\rceil.
\label{eq:reservation}
\end{equation}
With a smaller reservation, simultaneous residency of all communication CTAs
is not guaranteed, and additional queueing can delay the consumption of ready
splits. With a larger reservation, unchanged communication occupancy provides
no additional parallelism while further reducing compute capacity. Hence,
\(R(C)\) is the minimum spatial partition that realizes a given
communication-CTA parallelism, not an independent tuning parameter.

Register usage limits the current GatherRS kernel to one resident
communication CTA per A100 SM. Consequently,
\begin{equation}
o_{\mathrm{comm}}=1,\qquad R(C)=C.
\label{eq:comm-occupancy}
\end{equation}
Each concurrent communication CTA therefore requires a distinct SM residency
position. Based on its thread, register, and dynamic shared-memory
requirements, the selected \(128\times128\) GEMM2 kernel admits at most two
resident compute CTAs per SM, i.e., \(o_{\mathrm{comp}}=2\). After reserving
\(R(C)\) SM positions, the resident compute capacity is
\begin{equation}
P(C)=o_{\mathrm{comp}}\bigl(G-R(C)\bigr).
\label{eq:compute-capacity}
\end{equation}
For the current A100 kernel family, this reduces to
\begin{equation}
P(C)=2(G-C).
\label{eq:a100-compute-capacity}
\end{equation}
Adding one communication CTA therefore has two opposing effects: it adds one
communication worker but removes one compute SM, or two resident compute-CTA
slots. This dependency--residency coupling forms the resource constraint of
the pipeline model.

\subsection{Wave-Quantized Pipeline Model}

The partition must account for the actual work on both sides.
Table~\ref{tab:workload-notation} defines the workload and tile quantities
used below.

\begin{table}[!b]
\centering
\caption{Workload and tile notation.}
\label{tab:workload-notation}
\footnotesize
\setlength{\tabcolsep}{3pt}
\renewcommand{\arraystretch}{1.06}
\begin{tabular}{@{}lp{0.74\columnwidth}@{}}
\toprule
\textbf{Symbol} & \textbf{Definition} \\
\midrule
\(M\) & Global tokens before top-\(k\) expansion; \(M=BS\) in prefill. \\
\(B\) & Batch size. \\
\(D\) & Number of participating ranks (world size). \\
\(E\) & Number of local experts on the current rank. \\
\(M_e\) & Token--expert assignments received by local expert \(e\). \\
\(N\) & GEMM2 output dimension (Transformer hidden size). \\
\(K\) & Tensor-parallel local GEMM2 reduction dimension. \\
\(L\) & Number of splits in the GEMM2+GatherRS pipeline. \\
\(g\) & Number of input groups in grouped GEMM2. \\
\(T_M^{\mathrm{comp}},T_N^{\mathrm{comp}}\) & Compute-tile dimensions along \(M\) and \(N\). \\
\(T_M^{\mathrm{comm}},T_N^{\mathrm{comm}}\) & Communication-tile dimensions along tokens and hidden size. \\
\(W_{\mathrm{comp}}\) & Compute tiles in one split. \\
\(W_{\mathrm{comm}}\) & Communication tiles per split and rank. \\
\bottomrule
\end{tabular}
\end{table}

\textbf{Compute-CTA work.}
Expert \(e\) requires
\(\lceil M_e/T_M^{\mathrm{comp}}\rceil\) tiles along the token dimension.
The output width of one split is \(\lceil N/L\rceil\), requiring
\(\lceil\lceil N/L\rceil/T_N^{\mathrm{comp}}\rceil\) hidden-dimension tiles.
The compute-CTA work in one split is therefore
\begin{equation}
W_{\mathrm{comp}}
=
g
\left(
\sum_{e=1}^{E}
\left\lceil\frac{M_e}{T_M^{\mathrm{comp}}}\right\rceil
\right)
\left\lceil
\frac{\lceil N/L\rceil}{T_N^{\mathrm{comp}}}
\right\rceil .
\label{eq:compute-work}
\end{equation}
The routed counts \(M_e\) determine only the GEMM tiles implied by the
router; our method neither changes expert assignments nor reschedules work by
expert popularity. Given \(C\), the compute waves per split are
\begin{equation}
V_{\mathrm{comp}}(C)
=
\left\lceil\frac{W_{\mathrm{comp}}}{P(C)}\right\rceil.
\label{eq:compute-waves}
\end{equation}

\textbf{Communication-CTA work.}
Each rank ultimately owns approximately \(\lceil M/D\rceil\) original tokens.
The current GatherRS communication tile is
\(T_M^{\mathrm{comm}}\times T_N^{\mathrm{comm}}=128\times1024\).
Communication-CTA work per split is
\begin{equation}
W_{\mathrm{comm}}
=
\left\lceil
\frac{\lceil M/D\rceil}{T_M^{\mathrm{comm}}}
\right\rceil
\times
\left\lceil
\frac{\lceil N/L\rceil}{T_N^{\mathrm{comm}}}
\right\rceil .
\label{eq:communication-work}
\end{equation}
With \(C\) communication CTAs, the required waves are
\begin{equation}
V_{\mathrm{comm}}(C)
=
\left\lceil\frac{W_{\mathrm{comm}}}{C}\right\rceil.
\label{eq:communication-waves}
\end{equation}
These two wave expressions expose the central tradeoff. Increasing \(C\)
typically reduces \(V_{\mathrm{comm}}(C)\), but also reduces \(P(C)\) and can
increase \(V_{\mathrm{comp}}(C)\). Both expressions contain ceilings, so
performance changes through steps and plateaus rather than continuously.
Adding a CTA within the same wave plateau may provide no benefit; crossing a
communication-wave boundary can suddenly improve performance, while
simultaneously crossing a compute-wave boundary can abruptly degrade it.

\textbf{Split-level dependency.}
Wave counts describe how many execution rounds each role requires, but not
when those rounds may execute. GEMM2 and GatherRS are not independent:
communication can process only a split that compute CTAs have completed and
published, while the same communication CTAs sequentially serve multiple
splits. The latency of the GEMM2+GatherRS operator therefore depends on both work
quantities, compute publication times, and the communication queue.

The persistent GEMM visitor places the expert problems of each split
contiguously in a global tile sequence. With \(P(C)\) resident compute slots,
compute CTA \(b\) takes global tiles
\(b,b+P(C),b+2P(C),\ldots\). This assignment allows CTAs to execute some tiles
of later splits before every tile of an earlier split completes. The model
therefore derives each split's publication boundary from the global tile
sequence rather than treating splits as isolated compute phases.

We use one compute-CTA wave as the time unit. For split
\(s\in\{0,\ldots,L-1\}\), completing the first \(s+1\) splits requires
\((s+1)W_{\mathrm{comp}}\) compute tiles. Its readiness time is
\begin{equation}
A_s(C)
=
\left\lceil
\frac{(s+1)W_{\mathrm{comp}}}{P(C)}
\right\rceil.
\label{eq:split-arrival}
\end{equation}
Compute and communication waves have different service times. Let \(\rho\)
be the service-time ratio of one communication-CTA wave to one compute-CTA
wave. The time to process one split with \(C\) communication CTAs is
\begin{equation}
q(C)=\rho V_{\mathrm{comm}}(C).
\label{eq:comm-service}
\end{equation}
Communication for split \(s\) can begin only after its result is published at
\(A_s(C)\) and the communication CTAs finish the preceding split at
\(F_{s-1}(C)\). Defining \(F_{-1}(C)=0\), the completion time of split \(s\)
is
\begin{equation}
F_s(C)
=
\max\bigl(A_s(C),F_{s-1}(C)\bigr)+q(C).
\label{eq:split-finish}
\end{equation}
When \(A_s(C)>F_{s-1}(C)\), communication waits for computation; otherwise, a
ready split waits for communication service. The predicted latency of the
GEMM2+GatherRS operator is the completion time of the final split:
\begin{equation}
T(C)=F_{L-1}(C).
\label{eq:makespan}
\end{equation}
This recurrence places cross-split compute look-ahead, split-readiness
dependencies, and sequential communication service on one timeline,
capturing pipeline fill, steady state, and drain. Launch and fixed metadata
costs common to every \(C\) do not affect the partition choice, so the model
compares candidates using their relative \(T(C)\).

\textbf{Relative wave service rate.}
The parameter \(\rho\) maps compute and communication waves to a common time
scale. To separate communication service from split-readiness waiting, we
measure both service rates at a fixed \(C=16\), where communication CTAs can
continuously obtain ready work. Compute completes 25 cumulative waves in
497.36~\(\mu\)s, while communication continuously serves two splits in
400.19~\(\mu\)s, giving
\begin{equation}
\rho_{\mathrm{ref}}
=
\frac{400.19/2}{497.36/25}
\approx 10.05.
\label{eq:rho-reference}
\end{equation}
The reference rate is adjusted for GEMM2 compute intensity and communication
scale. A larger reduction dimension \(K\), or a larger GEMM2 tile, increases
the work in a compute wave and thereby decreases the time of a communication
wave relative to it; more ranks increase relative ring-communication cost.
At runtime, we use
\begin{equation}
\begin{split}
\rho
={}&
\rho_{\mathrm{ref}}
\frac{K_{\mathrm{ref}}}{K}
\frac{
T_{M,\mathrm{ref}}^{\mathrm{comp}}
T_{N,\mathrm{ref}}^{\mathrm{comp}}
}{
T_M^{\mathrm{comp}}T_N^{\mathrm{comp}}
}
\frac{D}{D_{\mathrm{ref}}},
\end{split}
\label{eq:rho-scaling}
\end{equation}
where
\begin{equation}
\begin{split}
K_{\mathrm{ref}}&=352,\\
T_{M,\mathrm{ref}}^{\mathrm{comp}}
&=T_{N,\mathrm{ref}}^{\mathrm{comp}}=128,\qquad
D_{\mathrm{ref}}=4.
\end{split}
\label{eq:rho-config}
\end{equation}
The conversion expresses \(A_s(C)\), \(q(C)\), and \(F_s(C)\) in compute-wave
units. The model can then compare the communication waves eliminated by
adding CTAs against the additional compute waves caused by their SM
reservation.

\subsection{Launch-Time Analytical Selection}

Before kernel launch, the resource manager receives the current workload
\(M,N,K,L,D\), host-side routed counts \(\{M_e\}\), the selected GEMM2
kernel's tile shape and occupancy, and the GPU SM count \(G\). For the current
A100 kernel family, the legal communication-CTA set is
\begin{equation}
\mathcal{C}=\{1,2,\ldots,C_{\max}\},\qquad C_{\max}=16.
\label{eq:c-domain}
\end{equation}
The upper bound comes from a one-time kernel-family saturation
characterization; it is not an A100 hardware limit. Section~\ref{Results}
validates this boundary.
The resource manager executes the procedure shown in Algorithm~\ref{alg:resource-manager} entirely on the CPU.


\begin{algorithm}[t]
\caption{CPU-side communication-grid selection}
\label{alg:resource-manager}
{\footnotesize
\begin{algorithmic}[1]
  \REQUIRE \(M,N,K,L,D,\{M_e\}\), kernel tile shape and occupancy, \(G\)
\STATE Compute \(W_{\mathrm{comp}}\) and \(W_{\mathrm{comm}}\)
\FOR{\(C=1\) to \(C_{\max}\)}
  \STATE \(R(C)\leftarrow\lceil C/o_{\mathrm{comm}}\rceil\)
  \STATE \(P(C)\leftarrow o_{\mathrm{comp}}(G-R(C))\)
  \STATE Compute all split-readiness times \(A_s(C)\)
  \STATE Compute \(V_{\mathrm{comm}}(C)\) and \(q(C)\)
  \STATE \(F_{-1}(C)\leftarrow 0\)
  \FOR{\(s=0\) to \(L-1\)}
    \STATE \(F_s(C)\leftarrow
      \max(A_s(C),F_{s-1}(C))+q(C)\)
  \ENDFOR
\ENDFOR
\STATE \(C^\ast\leftarrow\arg\min_{C\in\mathcal C}T(C)\)
\RETURN communication grid \(C^\ast\), communication reservation
\(R(C^\ast)\), compute capacity \(P(C^\ast)\)
\end{algorithmic}
}
\end{algorithm}

When multiple \(C\) values have the same wave-quantized \(T(C)\), the selector
uses the number of empty slots in the final compute wave,
\begin{equation}
\Delta_{\mathrm{comp}}(C)
=
V_{\mathrm{comp}}(C)P(C)-W_{\mathrm{comp}},
\label{eq:tie-breaker}
\end{equation}
as a tie breaker without additional fitted coefficients, preferring the
candidate with smaller \(\Delta_{\mathrm{comp}}(C)\). This rule reduces empty
slots in the final incomplete compute wave without using measurement noise to
distinguish candidates on the same predicted plateau.

The loop evaluates at most 16 \emph{analytical} candidates. It neither
launches nor times candidate GPU kernels and does not query a table of
measured per-shape winners. Its complexity is
\(O(E+L|\mathcal{C}|)\): the \(E\) term reduces routed expert tiles, while
\(L\) and \(|\mathcal{C}|\) are both small in the current implementation.

\subsection{System Integration and Correctness}

The resource manager sits between conventional kernel dispatch and the actual
launch. Existing dispatch still selects the GEMM2 compute kernel, after which
the manager reads its host-side tile shape and maximum resident blocks per SM.
The GPU SM count is cached during operator initialization. Tile shape and
occupancy are queried again only when dispatch selects a different GEMM2
kernel, avoiding repeated hardware-query calls on every MoE layer invocation.

At runtime, the manager preferentially uses host-side expert split counts
already available on the execution path to compute
\begin{equation}
\sum_e
\left\lceil
\frac{M_e}{T_M^{\mathrm{comp}}}
\right\rceil .
\label{eq:routed-tile-reduction}
\end{equation}
If routed counts exist only on the device, the system uses a normalized-load
estimate derived from the total assignments and expert count rather than
introducing a device-to-host copy or synchronization solely for scheduling.
The system launches \(C^\ast\) GatherRS communication CTAs, reserves
\(R(C^\ast)\) SM residency positions for communication, and limits the
remaining compute capacity to \(P(C^\ast)\). The modeled partition therefore
matches the actual kernel launch.

The baseline and our method reuse the same GEMM2 compute kernel, GatherRS
communication kernel, and readiness protocol. Input tensors, weights, router
assignments, GEMM tiles, communication volume, activations, and outputs remain
unchanged; only \(C\) and its resulting \(R(C)\)/\(P(C)\) partition differ.
An explicit user configuration retains highest priority,
and disabling analytical selection restores the upstream fixed configuration
and hot path. The measured improvements therefore come from rebalancing the
same compute and communication work, not from reducing model work or relaxing
correctness requirements.

The current implementation and service-rate parameters instantiate the model
for a single-node A100/NVLink kernel family. The same modeling method applies
to another fine-grained overlap pipeline when compute and communication CTAs
compete for GPU residency, a describable readiness DAG connects them, and
both workloads can be expressed as tile waves. Such an instantiation must,
however, rederive the kernel's \(o_{\mathrm{comp}}\),
\(o_{\mathrm{comm}}\), tile work, dependency recurrence, legal \(C\) range,
and service rate.

\section{Experimental Results}\label{Results}
\input{experimental_results.tex}

\section{Related Work}

Prior work spans distributed MoE execution, fine-grained computation-communication overlap, and GPU resource management.
These areas address different levels of the execution stack: distributed MoE systems organize operators and devices, overlap systems construct fine-grained dependency pipelines, and resource-management systems allocate execution capacity among concurrent GPU tasks.
Distributed MoE systems coordinate expert placement, token dispatch and combination, grouped expert computation, and inter-device communication.
Tutel adaptively selects parallelization strategies, All-to-All algorithms, and pipeline degrees~\cite{tutel}. PipeMoE models computation and communication to select a pipeline depth~\cite{pipemoe}.
ScheMoE schedules computation and communication over a task graph~\cite{schemoe}. FasterMoE uses expert shadowing to mitigate device-level load imbalance caused by dynamic routing~\cite{fastermoe}. 
These approaches primarily determine where a task runs, how it is partitioned, and when it starts, These approaches primarily determine where a task runs, how it is partitioned,
and when it starts, thereby exposing more opportunities for overlap. But they do not explicitly derive the resident compute/communication CTA ratio within a fine-grained dependency-coupled operator.

Fine-grained overlap efforts reduce synchronization at operator boundaries by exposing compute-to-communication dependencies at tile or split granularity. Recent works use readiness signaling, tile scheduling, work reordering, concurrent kernels, and thread-block specialization to allow communication to consume partial results before all computation completes~\cite{cui2026tile,eurosys26}.
COMET, the direct predecessor of this work, constructs such dependency pipelines through shared-tensor dependency analysis, tile rescheduling, horizontal fusion, and thread-block specialization~\cite{comet}.
FLUX~\cite{flux} decomposes computation and communication to tile granularity and combines kernel fusion with dependency synchronization to reduce exposed communication time. 
TileLink instead provides tile-centric primitives and a
compiler backend that generates overlap kernels with lower programming
effort~\cite{tilelink}. These systems establish execution mechanisms and
programming abstractions for fine-grained overlap. FlashDMoE further places dispatch, expert computation, and combine in a single persistent GPU kernel, using device-initiated communication to remove host orchestration from the critical path~\cite{flashdmoe}. 
Our work begins after such
dependency-coupled execution has been established and models and controls the
resource allocation between its compute and communication CTAs.

State-of-the-art approaches have also studied how GPU execution resources should be allocated and scheduled across concurrent operators, communication tasks, and tiled workloads. NanoFlow jointly selects operation order and GPU resource allocation for LLM serving through profiling- and interference-aware search~\cite{nanoflow}. Lagom uses a unified cost model and measurement-guided iterative search to configure multiple computation and communication operators in distributed LLM training ~\cite{lagom}. DeepEP V2 analytically determines SM and queue-pair requirements for expert-parallel communication from traffic and link characteristics~\cite{deepep}, while Stream-K analyzes the wave-quantization loss caused by unevenly distributed GEMM tiles~\cite{streamk}. These studies establish the importance of resource contention, communication provisioning, and discrete work waves. Our work focuses on a narrower dependency-coupled MoE operator, combining non-preemptive CTA residency, wave quantization, and fine-grained readiness dependencies to select the compute–communication allocation analytically at launch time without per-shape profiling.

\section{Conclusion}\label{Conclusion}

This paper presents a launch-time resource manager for fine-grained
computation--communication overlap in distributed MoE inference. Given the
workload, CTA residency constraints, and split-level dependencies, it selects
the communication-CTA count and the compute/communication resource partition
before kernel launch. We integrate the method into the COMET A100
implementation in FLUX.
Across three MoE models and three TP/EP configurations on four A100 GPUs, the
method achieves geometric-mean speedups over COMET of \(2.528\times\) for the
GEMM2+GatherRS operator, \(1.771\times\) for the complete post-router MoE
layer, and \(1.185\times\) for complete-model prefill. The corresponding
maximum speedups are \(4.218\times\), \(2.584\times\), and \(1.439\times\).
All 95\% confidence intervals for the operator and layer measurements remain
above \(1\times\).

Future work will evaluate the analytical model on newer GPU architectures and multi-node
interconnects, and extend resource selection to decode and other
computation--communication operators with fine-grained dependencies.

\section*{Acknowledgment}

Computational resources were provided by the National Academic Infrastructure for Supercomputing in Sweden (NAISS), funded by the Swedish Research Council. The computations were performed on the Alvis system hosted by Chalmers e-Commons at Chalmers University of Technology.

The authors used OpenAI Codex to assist with language editing.
The authors reviewed and revised the resulting text and take responsibility
for the manuscript.

\bibliographystyle{IEEEtran}
\bibliography{references}


\end{document}

%% file: background.tex
\section{Background and Problem Formulation}
\label{BK_PF}


This section provides the execution context needed to understand the resource
management problem studied in this work. We first describe the dataflow of a
distributed MoE layer, then explain how fine-grained computation--communication
overlap forms a dependency pipeline executed by compute and communication
CTAs, and finally formulate resource partitioning within this pipeline as the
problem addressed in this paper.

\subsection{Distributed MoE Execution}

Given a set of input hidden states, the router selects the top-\(k\) experts
for each token and produces token--expert assignments. The post-router MoE
layer studied in this work follows the dataflow of the existing fused
implementation and consists of three consecutive operations:
\begin{enumerate}
\item \textbf{AllGather+GEMM1.} AllGather collects the participating input
segments. As the required data arrives, grouped GEMM1 performs the gate and up
projections.
\item \textbf{Activation.} A gated activation combines the two projections to
form the intermediate representation for each expert.
\item \textbf{GEMM2+GatherRS.} Grouped GEMM2 projects the expert intermediate
representations back to the hidden dimension. GatherRS aggregates the
top-\(k\) expert contributions, performs ReduceScatter, and writes the token
outputs for each rank.
\end{enumerate}

\begin{figure}[!t]
\centering
\includegraphics[width=\columnwidth]{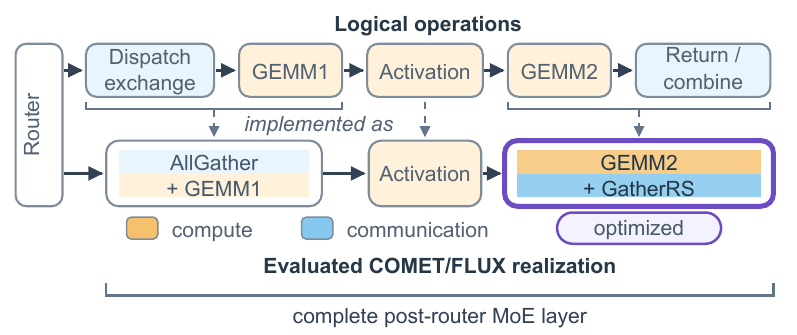}
\caption{Post-router MoE path and its evaluated COMET/FLUX implementation.
The highlighted GEMM2+GatherRS operator is the optimization target.}
\label{fig:post-router-moe-path}
\end{figure}

Fig.~\ref{fig:post-router-moe-path} relates the logical post-router operator
sequence to the evaluated COMET/FLUX realization.
Our resource decision acts on the final GEMM2+GatherRS operator. To avoid ambiguity, we
use \emph{GEMM2+GatherRS operator} for this joint computation--communication
operator, \emph{complete post-router MoE layer} for the path from AllGather+GEMM1
through the GatherRS output writeback, and \emph{complete-model prefill} for a
complete model prefill forward pass, including attention, routing, MoE blocks,
GPU-resident KV-cache updates, and the last-token LM head. These three
boundaries isolate the direct effect of the resource decision, its propagation
through the MoE layer, and its model-level impact, respectively.

\subsection{Fine-Grained Overlap and GPU Residency}

A CUDA thread block, also called a cooperative thread array (CTA), comprises
multiple warps and is scheduled on one SM. While
resident, a CTA occupies threads, registers, and shared memory; once admitted
in ordinary kernel execution, it normally retains these resources until
completion. These resource requirements jointly determine how many CTAs can
reside on an SM. A CTA is therefore neither a single warp nor an SM: one SM
may host multiple CTAs when resources permit, whereas one CTA is not split
across multiple SMs.

\begin{figure}[!t]
\centering
\includegraphics[width=\columnwidth]{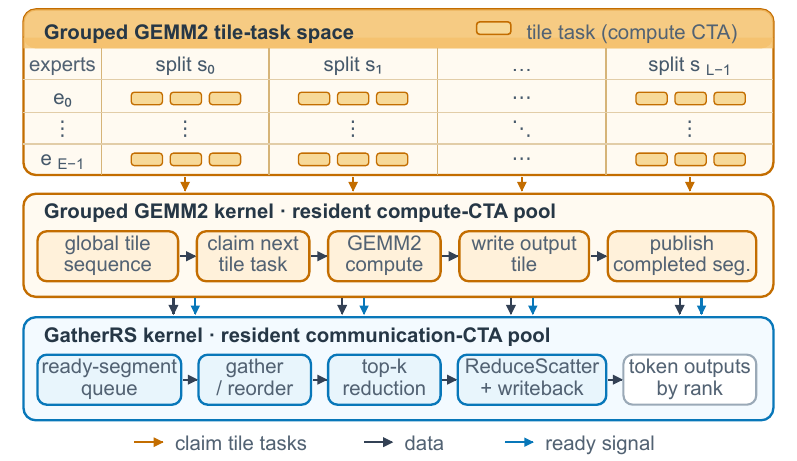}
\caption{Split-level execution of the GEMM2+GatherRS operator. Compute CTAs
publish a segment after all tiles in the corresponding split complete;
communication CTAs consume ready segments.}
\label{fig:gemm2-gatherrs-cta-pipeline}
\end{figure}

In the fine-grained overlap path studied here, \emph{compute CTAs} repeatedly
claim and execute grouped-GEMM2 tile tasks, as shown in
Fig.~\ref{fig:gemm2-gatherrs-cta-pipeline}. A segment becomes ready only after
every tile task in its corresponding split finishes; the compute side then
publishes its data and readiness signal. \emph{Communication CTAs} consume
these ready segments to aggregate expert outputs, reduce them, and write back
the corresponding result. Communication can therefore begin before the
complete GEMM2 finishes, forming a dependency pipeline in which computation
publishes completed segments and communication consumes them. A communication
CTA can perform useful work only after its input becomes ready, and the same
communication CTAs service multiple splits in sequence. Concurrent residency
alone therefore does not guarantee balanced progress.

Both CTA classes share finite GPU residency. Increasing the number of
communication CTAs can process more ready communication tiles in parallel and
reduce communication waves, but it also occupies SM resources that would
otherwise be available to compute CTAs. Compute parallelism consequently
decreases and later split-release times may be delayed. Conversely, too few
communication CTAs cause ready splits to queue and leave a communication tail
on the completion path of the GEMM2+GatherRS operator. Both workloads execute in integer
CTA waves: completion time changes materially only when a resource adjustment
adds or removes a wave. Performance as a function of the resource partition
therefore exhibits plateaus and steps rather than smooth linear scaling.

Existing fine-grained overlap mechanisms already specify tile decomposition,
split-level readiness, and the concurrent execution of the two CTA classes.
Our work instead asks how, once this mechanism is fixed, a particular
invocation should partition residency between the two classes so that exposed
concurrency becomes effective overlap.

\subsection{Optimization Scope and Problem Formulation}

We study the public A100 implementation of COMET in the FLUX
codebase~\cite{comet,flux}. This path fixes the compute and communication
kernels, their tile- and split-level readiness protocol, and their dataflow.
The public A100 path uses a fixed communication-CTA count \(C=3\), which can
be overridden manually but is not selected analytically for each invocation.

Let \(\mathcal{W}\) describe an invocation's workload, including its matrix
dimensions, split count, and routed expert-tile counts; let \(\mathcal{K}\)
describe the tile shape and resource footprint of the kernels selected by
dispatch; let \(\mathcal{H}\) describe the GPU SM count and residency
constraints; and let \(\mathcal{G}_{\mathrm{dep}}\) denote the split-level
dependency graph between compute results and communication tasks. The only
decision variable is the communication-CTA count \(C\) for the current launch.
Given \(C\), the communication parallelism, its required residency reservation,
and the remaining compute-CTA capacity are jointly determined.

To isolate the effect of resource partitioning, comparisons across \(C\) hold
\(\mathcal{W}\), \(\mathcal{K}\), and \(\mathcal{G}_{\mathrm{dep}}\) fixed.
The same inputs and router assignments produce the same GEMM tiles, perform the
same arithmetic work, and communicate the same data. The resource manager
changes only the communication-CTA grid and the resulting compute-CTA
reservation. The problem therefore does not reroute tokens or change a kernel
algorithm; it changes how the same work and dependencies occupy GPU execution
resources.

Let
\(T(C\mid\mathcal{W},\mathcal{K},\mathcal{H},
\mathcal{G}_{\mathrm{dep}})\) denote the completion time from the launch of the
first compute CTA to the writeback of the final communication result under a
given partition. The legal set \(\mathcal{C}\) contains integer CTA counts that
reserve sufficient communication residency while retaining positive compute
capacity. Before the actual launch, we seek
\begin{equation}
C^\ast =
\underset{C\in\mathcal{C}}{\arg\min}\;
T
\left(C\mid\mathcal{W},\mathcal{K},\mathcal{H},
\mathcal{G}_{\mathrm{dep}}\right).
\label{eq:resource-partition-objective}
\end{equation}

This objective imposes three requirements: communication CTAs must obtain
enough residency to consume ready splits, compute CTAs must retain enough
capacity to produce subsequent results, and the selector cannot learn the
answer by executing and timing candidate kernels. The core problem is therefore
to use the workload, kernel residency, and readiness dependencies available
before launch to predict the two discrete wave sequences analytically and
select the communication-CTA count that minimizes GEMM2+GatherRS completion time.
The following section derives the progress-preserving reservation, the compute
and communication wave counts, the split-readiness recurrence, and the
corresponding launch-time selection algorithm.

%% file: experimental_results.tex
\begin{figure*}[!t]
\centering
\includegraphics[width=0.78\textwidth]{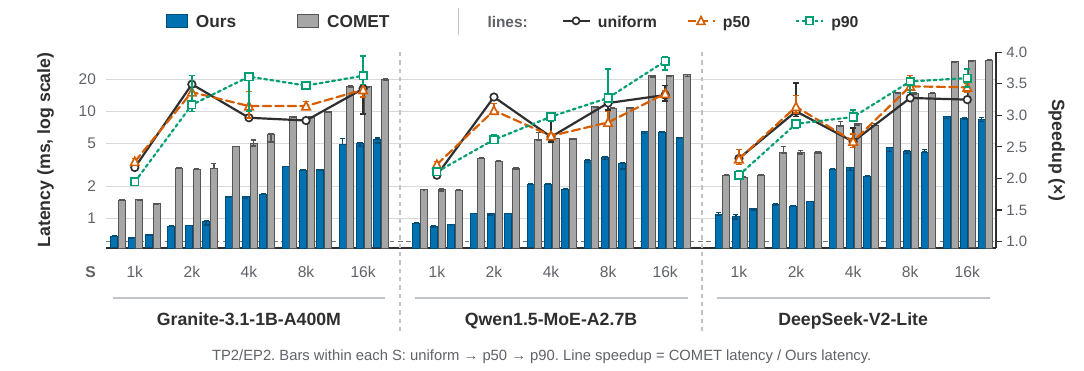}
\caption{Latency and speedup of the GEMM2+GatherRS operator versus sequence
length~\(S\) under uniform and real-router p50/p90 workloads with TP=2 and
EP=2.}
\label{fig:gemm2-gatherrs-stage}
\end{figure*}

\subsection{Experimental Setup}
\subsubsection{Hardware and Software}
All experiments run on the same compute node with four NVIDIA A100-SXM4-40GB GPUs connected by NVLink. They use BF16 precision and evaluate three parallelism configurations: \(\mathrm{TP}=4/\mathrm{EP}=1\), \(\mathrm{TP}=2/\mathrm{EP}=2\), and \(\mathrm{TP}=1/\mathrm{EP}=4\), where TP and EP denote the tensor-parallel and expert-parallel degrees and \(\mathrm{TP}\times\mathrm{EP}=4\) in every configuration.
We use PyTorch 2.7.1, CUDA 12.6, NCCL 2.26.2, CUTLASS commit \texttt{df8a550}, and COMET A100 V2 commit \texttt{19831ca}; Granite uses one pipeline split, and the other models use two.

\subsubsection{Models and Workloads}
Table~\ref{tab:model-configs} summarizes the three evaluated models. All experiments use batch size \(B=4\) and sequence length \(S\in\{1024,2048,4096,8192,16384\}\).
GEMM2+GatherRS operator and MoE layer experiments
use uniform routing with equal expert probabilities and real traces produced by the native model routers on LMSYS-Chat-1M~\cite{lmsyschat1m}.
For each model and sequence length, we select the actual assignment tensors nearest the empirical p50 and p90 of \(R_{\mathrm{expert}}=\max_e L_e/(Mk/E)\). 
COMET and our method replay the same selected tensor. 
Complete-model prefill instead uses the natural per-layer routing of the forward pass.
The three models, three routing workloads, and five input lengths form 45
evaluated model--routing--sequence configurations for each TP/EP
configuration.

\begin{table}[!b]
\centering
\caption{MoE model configurations. \(H\) and \(F\) denote the hidden and
expert-FFN sizes, respectively.}
\label{tab:model-configs}
\footnotesize
\setlength{\tabcolsep}{2.5pt}
\begin{tabular}{@{}lcccc@{}}
\toprule
\textbf{Model} & \(\boldsymbol{H}\) & \(\boldsymbol{F}\) &
\textbf{Routed experts} & \textbf{Top-\(\boldsymbol{k}\)} \\
\midrule
Granite-3.1-1B-A400M~\cite{granite31a400m} & 1024 & 512 & 32 & 8 \\
Qwen1.5-MoE-A2.7B~\cite{qwen15moe} & 2048 & 1408 & 60 & 4 \\
DeepSeek-V2-Lite~\cite{deepseekv2} & 2048 & 1408 & 64 & 6 \\
\bottomrule
\end{tabular}
\end{table}


\subsubsection{Measurement Methodology and Correctness}
After 20 warm-up iterations, operator and layer measurements collect 60 ABBA-interleaved samples per policy, while complete-model and multi-backend measurements collect 20 per backend. We report median maximum-rank latency and 95\% confidence intervals from 20,000 block-stratified bootstrap replicates.
COMET and ours pass four-rank \texttt{allclose} and produce identical complete-model outputs.
Against Hugging Face references, all models preserve last-token top-1 predictions, and Qwen and DeepSeek-V2-Lite additionally pass elementwise \texttt{allclose}.

\subsubsection{Baselines}
The primary baseline is the public COMET A100 path described in Section~\ref{BK_PF}. \textbf{Ours} changes only \(C\) and the resulting residency partition while retaining the same kernels, assignments, arithmetic work, and communication volume.
For complete-model prefill, we additionally compare with \textbf{Megatron core-TE} and \textbf{FastMoE TP+NCCL}~\cite{megatroncore,transformerengine,fastmoe}. Megatron core-TE uses Transformer Engine GroupedLinear with NCCL AllGather and
ReduceScatter, whereas FastMoE TP+NCCL uses its tensor-parallel expert path
with the same NCCL collectives. 

Due to page limitation, we present detailed results for the representative balanced configuration when $TP=2, EP=2$ and summarize the other two configurations in Fig.~\ref{fig:cross-parallelism}.

\subsection{Predictor Accuracy and Runtime Overhead} 

For each model's real-p90 TP=4/EP=1 workload at all five sequence lengths, we
obtain the measured oracle by sweeping \(C=1,\ldots,16\) and selecting the
lowest median latency. We report
\(\mathrm{regret}=(T_{\mathrm{predicted}}-T_{\mathrm{oracle}})
/T_{\mathrm{oracle}}\). Across the 15 workloads, the mean and maximum regrets
are 3.22\% and 10.21\%, respectively (Table~\ref{tab:predictor-accuracy}).

\begin{table}[!b]
\centering
\caption{Predictor regret relative to the measured oracle.}
\label{tab:predictor-accuracy}
\footnotesize
\setlength{\tabcolsep}{3pt}
\begin{tabular}{@{}lcc@{}}
\toprule
\textbf{Evaluation matrix} & \textbf{Mean} & \textbf{Maximum} \\
\midrule
Granite real-p90, five shapes & 0.91\% & 4.13\% \\
Qwen real-p90, five shapes & 7.14\% & 10.21\% \\
DeepSeek-V2-Lite real-p90, five shapes & 1.60\% & 6.36\% \\
\midrule
All 15 workloads & 3.22\% & 10.21\% \\
\bottomrule
\end{tabular}
\end{table}

\begin{figure*}[!t]
\centering
\includegraphics[width=0.78\textwidth]{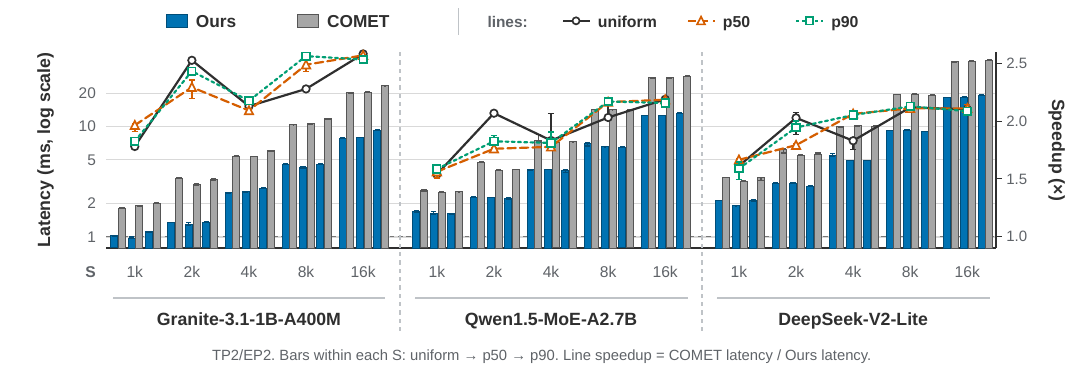}
\caption{Complete post-router MoE layer latency and speedup versus sequence
length~\(S\) under uniform and real-router p50/p90 workloads with TP=2 and
EP=2.}
\label{fig:complete-post-router-moe-layer}
\end{figure*}

The legal domain \(C\leq16\) is frozen by a one-time A100 kernel-family
characterization; an extended sweep to \(C=32\) on a communication-intensive
boundary workload found no out-of-domain improvement. The mean standalone
solver overhead is 0.157~\(\mu\)s. All subsequent measurements include
routed-count extraction, metadata handling, and operator dispatch.

\subsection{GEMM2+GatherRS Operator Evaluation}
Fig.~\ref{fig:gemm2-gatherrs-stage} reports the absolute latency of the
\texttt{GEMM2+GatherRS} operator
and its speedup over upstream COMET.
The five-length geometric-mean speedups for uniform, real-p50, and
real-p90 are \(2.951\times\), \(3.034\times\), and \(3.086\times\) for
Granite; \(2.856\times\), \(2.813\times\), and \(2.905\times\) for Qwen; and
\(2.870\times\), \(2.948\times\), and \(2.945\times\) for DeepSeek-V2-Lite.
The 95\% confidence interval is above \(1\times\) at every point.
The gains come from resource rebalancing rather than less communication or
GEMM work. For Granite real-p50 at \(S=2048\), for example, latency of the
GEMM2+GatherRS operator decreases from 2.8824 to
0.8564~ms, a \(3.366\times\) speedup. 
The best TP=2/EP=2 result is Qwen
real-p90 at \(S=16384\), where latency decreases from 22.0088 to
5.7067~ms, or \(3.857\times\). As \(S\) and the model shape change, the
predictor adjusts \(C\) from the compute and communication wave counts, avoiding
the persistent mismatch between communication progress and GEMM capacity
created by COMET's static resource partition.

Similar aggregate gains on uniform and real traces show that the predictor
remains effective across routing distributions, despite changes in the
compute-tile count caused by expert-boundary rounding.

\begin{figure*}[!t]
\centering
\includegraphics[width=0.78\textwidth]{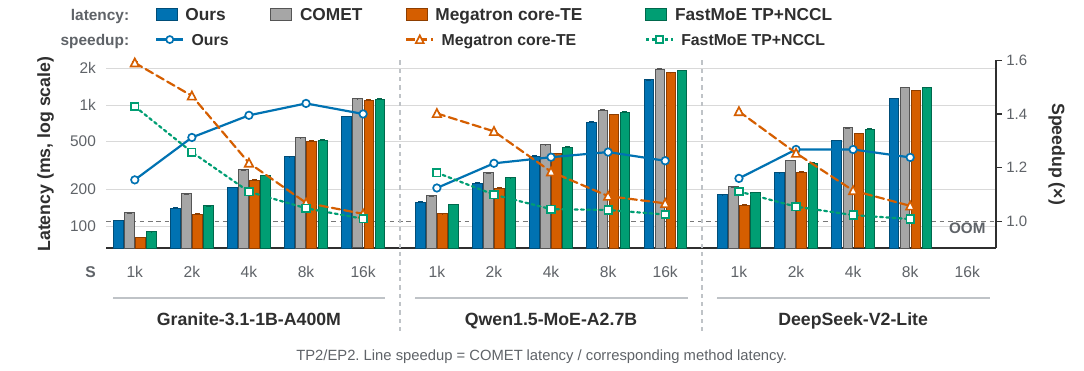}
\caption{Complete-model prefill latency and speedup versus sequence
length~\(S\) across four distributed MoE implementations with TP=2 and EP=2.}
\label{fig:complete-model-prefill}
\end{figure*}

\subsection{Complete Post-Router MoE Layer Evaluation}

Fig.~\ref{fig:complete-post-router-moe-layer} reports complete post-router MoE layer latency and speedup for TP=2 and EP=2. 
Because our method leaves AllGather+GEMM1 and the activation unchanged, 
the layer-level speedups are smaller than the operator-level results.
The five-length geometric-mean
speedups for uniform, real-p50, and real-p90 are \(2.241\times\),
\(2.271\times\), and \(2.290\times\) for Granite; \(1.918\times\),
\(1.875\times\), and \(1.897\times\) for Qwen; and \(1.929\times\),
\(1.939\times\), and \(1.950\times\) for DeepSeek-V2-Lite. 
For example,
Granite real-p90 at \(S=4096\) decreases from 6.0426 to 2.7754~ms, a
\(2.177\times\) speedup; DeepSeek-V2-Lite real-p90 at \(S=2048\) decreases
from 5.6394 to 2.8993~ms, a \(1.945\times\) speedup.

At every evaluated point with \(B=4\) and \(S\geq1024\), the 95\% confidence
interval for the complete post-router MoE layer remains above \(1\times\).

\subsection{Complete Model Prefill Evaluation}
\begin{figure}[!t]
\centering
\includegraphics[width=\columnwidth]{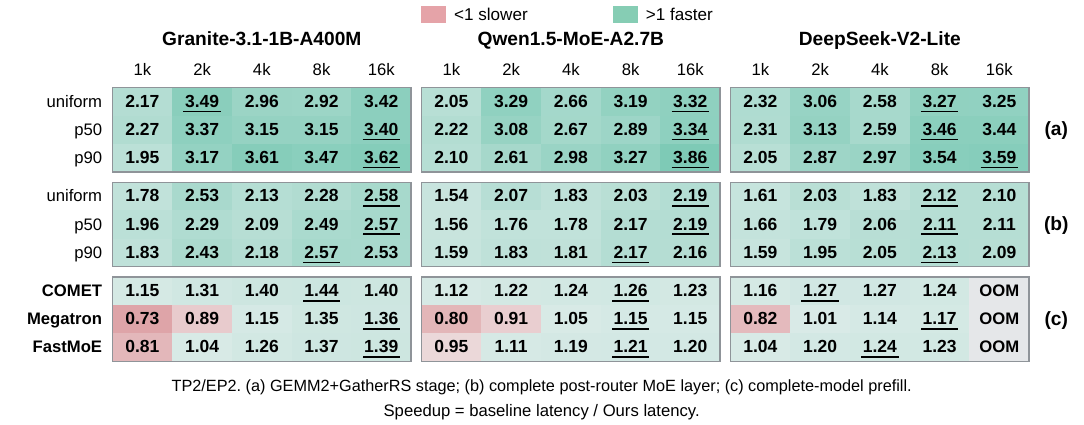}
\caption{Pointwise speedup with TP=2/EP=2 at the GEMM2+GatherRS operator,
complete post-router MoE layer, and complete-model prefill levels. Values are
baseline latency divided by our latency; green and red indicate speedup and
slowdown. Underlines mark row
maxima, and OOM indicates an infeasible run. Megatron and FastMoE denote
Megatron core-TE and FastMoE TP+NCCL.}
\label{fig:speedup-summary}
\end{figure}

Fig.~\ref{fig:complete-model-prefill} reports complete-model prefill latency,
and Table~\ref{tab:complete-model-main} gives all paired results for COMET
and our method 
for TP=2 and EP=2. 
Granite and Qwen cover five sequence lengths
and obtain geometric-mean speedups of \(1.336\times\) and \(1.211\times\),
respectively. DeepSeek-V2-Lite covers four feasible lengths and has a
geometric-mean speedup of \(1.232\times\). The best point is Granite at
\(S=8192\), where latency
decreases from 539.44 to 374.76~ms, a \(1.439\times\) speedup. As the input
grows, the time saved in the GEMM2+GatherRS operator accumulates across MoE
blocks, offsetting launch-time metadata and dispatch costs and surviving the
unchanged attention, normalization, router, KV-cache update, and LM-head
work. 
Fig.~\ref{fig:speedup-summary} consolidates the
pointwise
speedups across all three measurement levels.

\begin{table}[!b]
\centering
\caption{Complete-model prefill results against COMET with TP=2 and EP=2.}
\label{tab:complete-model-main}
\footnotesize
\setlength{\tabcolsep}{2.2pt}
\begin{tabular}{@{}lrrrr@{}}
\toprule
\textbf{Model} & \(\boldsymbol{S}\) & \textbf{COMET (ms)} &
\textbf{Ours (ms)} & \textbf{Speedup} \\
\midrule
Granite & 1024  & 128.78  & 111.59  & \(1.154\times\) \\
Granite & 2048  & 185.19  & 141.11  & \(1.312\times\) \\
Granite & 4096  & 292.25  & 209.49  & \(1.395\times\) \\
Granite & 8192  & 539.44  & 374.76  & \(1.439\times\) \\
Granite & 16384 & 1130.41 & 807.37  & \(1.400\times\) \\
\midrule
Qwen & 1024  & 178.32  & 158.68  & \(1.124\times\) \\
Qwen & 2048  & 276.08  & 227.05  & \(1.216\times\) \\
Qwen & 4096  & 471.08  & 380.32  & \(1.239\times\) \\
Qwen & 8192  & 911.45  & 724.61  & \(1.258\times\) \\
Qwen & 16384 & 1982.85 & 1618.15 & \(1.225\times\) \\
\midrule
DeepSeek-V2-Lite & 1024 & 211.28  & 182.21  & \(1.160\times\) \\
DeepSeek-V2-Lite & 2048 & 350.85  & 276.77  & \(1.268\times\) \\
DeepSeek-V2-Lite & 4096 & 648.49  & 511.66  & \(1.267\times\) \\
DeepSeek-V2-Lite & 8192 & 1404.36 & 1134.45 & \(1.238\times\) \\
\bottomrule
\end{tabular}
\end{table}


Fig.~\ref{fig:complete-model-prefill} also compares four implementations under
identical model inputs and measurement boundaries.
Table~\ref{tab:backend-speedups} reports the geometric-mean speedup of our
method over each implementation across lengths; DeepSeek-V2-Lite uses its
four feasible lengths, while the other models use five.
These means do not imply that our method is always fastest at the shortest
input. At \(S=1024\), Megatron core-TE has the lowest latency on all three models;
at Qwen \(S=2048\), its 206.78~ms is also faster than our 227.05~ms. The
crossover appears as tile-wave counts grow:
our method leads Granite and Qwen from \(S=4096\) onward and
DeepSeek-V2-Lite from \(S=2048\) onward. It is the fastest of all four
implementations at every feasible TP=2/EP=2 point with \(S\geq4096\). This
trend matches the design target: fixed-partition costs become more exposed
when enough token tiles form multiple compute and communication waves, whereas
short workloads are more readily dominated by launches, collectives, and
unchanged dense operations.

DeepSeek-V2-Lite complete-model prefill at \(B=4,S=16384\) exceeds the memory
capacity of a 40-GB A100. We retain the Out-of-Memory (OOM) result rather than reducing the
batch size or sequence length or changing the KV-cache boundary. The
GEMM2+GatherRS operator and complete post-router MoE layer at the same shape do
not retain the full model state, so they remain feasible and are included in
the first two result sets.
Accordingly, these results apply to the complete-model prefill boundary
defined above, not to complete online-serving or decode throughput.

\begin{table}[!t]
\centering
\caption{Geometric-mean complete-model prefill speedup of ours over each
backend with TP=2 and EP=2.}
\label{tab:backend-speedups}
\footnotesize
\setlength{\tabcolsep}{2.5pt}
\begin{tabular}{@{}lrrr@{}}
\toprule
\textbf{Model} & \textbf{COMET} & \textbf{Megatron core-TE} &
\textbf{FastMoE TP+NCCL} \\
\midrule
Granite & \(1.336\times\) & \(1.064\times\) & \(1.151\times\) \\
Qwen & \(1.211\times\) & \(1.002\times\) & \(1.125\times\) \\
DeepSeek-V2-Lite & \(1.232\times\) & \(1.026\times\) & \(1.176\times\) \\
\bottomrule
\end{tabular}
\end{table}

\begin{figure}[!t]
\centering
\includegraphics[width=\columnwidth]{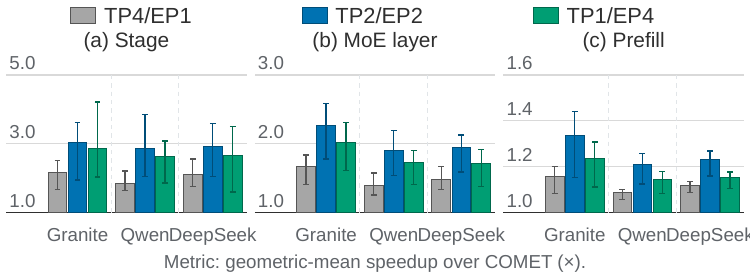}
\caption{Speedup over COMET across the three four-GPU TP/EP configurations.
Bars show geometric means and whiskers show pointwise ranges:
(a) GEMM2+GatherRS operator; (b) complete post-router MoE layer;
(c) complete-model prefill.}
\label{fig:cross-parallelism}
\end{figure}

\subsection{Cross-Configuration Summary}

Fig.~\ref{fig:cross-parallelism} summarizes the speedups over COMET across
all three TP/EP configurations. Every configuration maintains a pointwise
minimum above \(1\times\) at all three
measurement levels. The lowest observed speedup is \(1.058\times\), obtained
for Qwen complete-model prefill with TP=4/EP=1 at \(S=1024\).
TP=2/EP=2 achieves the largest geometric-mean speedup at every
measurement level, followed by TP=1/EP=4 and TP=4/EP=1.
Across all configurations, the geometric-mean speedups over COMET are \(2.528\times\) at the GEMM2+GatherRS operator, \(1.771\times\) at the complete post-router MoE layer, and \(1.185\times\) for complete-model prefill. The corresponding maximum speedups are \(4.218\times\), \(2.584\times\), and \(1.439\times\), respectively.